\documentclass[twocolumn,showpacs,aps,prl,superscriptaddress]{revtex4}

\usepackage{graphicx}
\usepackage{dcolumn}
\usepackage{epsfig}
\RequirePackage{xspace}

\usepackage{relsize}
\def\babar{\mbox{\slshape B\kern-0.1em{\smaller A}\kern-0.1em
    B\kern-0.1em{\smaller A\kern-0.2em R}}}

\def\epem       {\ensuremath{e^+e^-}\xspace}

\def\qqbar {\ensuremath{q\overline q}\xspace}

\def\pip   {\ensuremath{\pi^+}\xspace}
\def\pim   {\ensuremath{\pi^-}\xspace}

\def\Kbar  {\kern 0.2em\overline{\kern -0.2em K}{}\xspace}

\def\Kz    {\ensuremath{K^0}\xspace}
\def\Kzb   {\ensuremath{\Kbar^0}\xspace}
\def\KzKzb {\ensuremath{\Kz \kern -0.16em \Kzb}\xspace}
\def\Kp    {\ensuremath{K^+}\xspace}
\def\Km    {\ensuremath{K^-}\xspace}

\def\KpKm  {\ensuremath{\Kp \kern -0.16em \Km}\xspace}
\def\KS    {\ensuremath{K^0_{\scriptscriptstyle S}}\xspace}

\def\Kstarzb {\ensuremath{\Kbar^{*0}}\xspace}

\def\Kstarp  {\ensuremath{K^{*+}}\xspace}

\def\Dbar    {\kern 0.2em\overline{\kern -0.2em D}{}\xspace}

\def\Dz      {\ensuremath{D^0}\xspace}
\def\Dzb     {\ensuremath{\Dbar^0}\xspace}
\def\DzDzb   {\ensuremath{\Dz {\kern -0.16em \Dzb}}\xspace}
\def\Dp      {\ensuremath{D^+}\xspace}
\def\Dm      {\ensuremath{D^-}\xspace}

\def\DpDm    {\ensuremath{\Dp {\kern -0.16em \Dm}}\xspace}

\def\Ds      {\ensuremath{D^+_s}\xspace}

\def\Dss     {\ensuremath{D^{*+}_s}\xspace}
\def\Dsos    {\ensuremath{D^{(*)+}_s}\xspace}

\def\B       {\ensuremath{B}\xspace}
\def\Bbar    {\kern 0.18em\overline{\kern -0.18em B}{}\xspace}

\def\BB      {\ensuremath{B\Bbar}\xspace} 
\def\Bz      {\ensuremath{B^0}\xspace}
\def\Bzb     {\ensuremath{\Bbar^0}\xspace}
\def\BzBzb   {\ensuremath{\Bz {\kern -0.16em \Bzb}}\xspace}
\def\Bu      {\ensuremath{B^+}\xspace}
\def\Bub     {\ensuremath{B^-}\xspace}

\def\BpBm    {\ensuremath{\Bu {\kern -0.16em \Bub}}\xspace}

\mathchardef\Upsilon="7107
\def\Y#1S{\ensuremath{\Upsilon{(#1S)}}\xspace}

\def\FourS {\Y4S}

\mathchardef\Deltares="7101
\mathchardef\Xi="7104
\mathchardef\Lambda="7103
\mathchardef\Sigma="7106
\mathchardef\Omega="710A

\def\Deltabar{\kern 0.25em\overline{\kern -0.25em \Deltares}{}\xspace}
\def\Lbar{\kern 0.2em\overline{\kern -0.2em\Lambda\kern 0.05em}\kern-0.05em{}\xspace}
\def\Sigbar{\kern 0.2em\overline{\kern -0.2em \Sigma}{}\xspace}
\def\Xibar{\kern 0.2em\overline{\kern -0.2em \Xi}{}\xspace}
\def\Obar{\kern 0.2em\overline{\kern -0.2em \Omega}{}\xspace}
\def\Nbar{\kern 0.2em\overline{\kern -0.2em N}{}\xspace}
\def\Xb{\kern 0.2em\overline{\kern -0.2em X}{}\xspace}

\def\BR         {{\ensuremath{\cal B}\xspace}}

\def\mes        {\mbox{$m_{\rm ES}$}\xspace}

\newcommand{\tev}{\ensuremath{\mathrm{\,Te\kern -0.1em V}}\xspace}
\newcommand{\gev}{\ensuremath{\mathrm{\,Ge\kern -0.1em V}}\xspace}
\newcommand{\mev}{\ensuremath{\mathrm{\,Me\kern -0.1em V}}\xspace}
\newcommand{\kev}{\ensuremath{\mathrm{\,ke\kern -0.1em V}}\xspace}
\newcommand{\ev}{\ensuremath{\mathrm{\,e\kern -0.1em V}}\xspace}
\newcommand{\gevc}{\ensuremath{{\mathrm{\,Ge\kern -0.1em V\!/}c}}\xspace}
\newcommand{\mevc}{\ensuremath{{\mathrm{\,Me\kern -0.1em V\!/}c}}\xspace}
\newcommand{\gevcc}{\ensuremath{{\mathrm{\,Ge\kern -0.1em V\!/}c^2}}\xspace}
\newcommand{\mevcc}{\ensuremath{{\mathrm{\,Me\kern -0.1em V\!/}c^2}}\xspace}

\def\mus  {\ensuremath{\rm \,\mus}\xspace}

\def\mus        {\ensuremath{\,\mu{\rm s}}\xspace}    

\def\to                 {\ensuremath{\rightarrow}\xspace}

\def\pep2{PEP-II}

\def\gsim{{~\raise.15em\hbox{$>$}\kern-.85em
          \lower.35em\hbox{$\sim$}~}\xspace}
\def\lsim{{~\raise.15em\hbox{$<$}\kern-.85em
          \lower.35em\hbox{$\sim$}~}\xspace}

\def\CP                {\ensuremath{C\!P}\xspace}

\xspace

\newcommand{\jprlBase}       {Phys.\ Rev.\ Lett.\xspace}
\newcommand{\jprBase}        {Phys.\ Rev.\xspace}
\newcommand{\jplBase}        {Phys.\ Lett.\xspace}

\newcommand{\nimBaseC}       {Nucl.\ Instr.\ and Methods\xspace}

\newcommand{\zpBase}         {Z.\ Phys.\xspace}

\newcommand{\nim}       [1]  {\nimBaseC~{\bf #1}}
\newcommand{\plb}       [1]  {\jplBase\ B~{\bf #1}}

\newcommand{\jprl}      [1]  {\jprlBase\ {\bf #1}}
\newcommand{\pr}        [1]  {\jprBase\ {\bf #1}}
\newcommand{\jprd}      [1]  {\jprBase\ D~{\bf #1}}

\newcommand{\progtp}    [1]  {{Prog.\ Th.\ Phys.\ {\bf #1}}}

\newcommand{\zpc}       [1]  {\zpBase\ C~{\bf #1}}

\def\jetset74   {\mbox{\tt Jetset \hspace{-0.5em}7.\hspace{-0.2em}4}\xspace}

\def\bdspi    {\ensuremath {\Bz{\to}D_s^{+}\pi^-}}
\def\bdsspi    {\ensuremath {\Bz{\to}D_s^{*+}\pi^-}}
\def\bdsk    {\ensuremath {\Bz{\to}D_s^- K^+}}

\def\bdssk    {\ensuremath {\Bz{\to}D_s^{*-}K^+}}
\def\bdsordsspi   {\ensuremath {\Bz{\to}D_s^{(*)+}\pi^-}}
\def\bdsordssk   {\ensuremath {\Bz{\to}D_s^{(*)-}K^+}}
\def\bbardsordssk   {\ensuremath {\Bzb{\to}D_s^{(*)+}K^-}}

\def\sin2bg       {\ensuremath {\sin (2\beta+\gamma)}}
\def\btou       {\ensuremath {b\rightarrow u}}

\def\brdsShort {\ensuremath { 3.2 \pm 0.9\, ({\rm stat. }) }}

\def\brdskShort {\ensuremath{ 3.2 \pm 1.0\, ({\rm stat. }) }}

\def\brds {\BR(\bdspi)~=~(\brdsShort $\pm 1.0\, ({ \rm syst. })$) \ensuremath{\times 10^{-5}}}

\def\brdsk {\BR(\bdsk)~=~(\brdskShort$\pm 1.0\, ({ \rm syst. })$) \ensuremath{\times 10^{-5}}}

\def\brdsslim {\BR(\bdsspi)\ensuremath{<4.1{\times}10^{-5}}}
\def\brdssklim {\BR(\bdssk)\ensuremath{<2.5{\times}10^{-5}}}

\def\lumi    {\ensuremath {84\ \rm million}}
\def\mds     {\ensuremath {M_{D_s}^{\rm{cand}}}}

\def\cth     {\ensuremath {\cos{\theta_{T}}}}
\def\cthel     {\ensuremath {\cos{\theta_{H}}}}
\def\ctwothel     {\ensuremath {\cos^2{\theta_{H}}}}

\def\de        {\ensuremath {\Delta E}}

\def\myprl  #1 #2 #3 {\jprl{#1},\ #2 (#3)}
\def\myplb  #1 #2 #3 {\plb{#1},\ #2 (#3)}
\def\myprd  #1 #2 #3 {\jprd{#1},\ #2 (#3)}
\def\mynim  #1 #2 #3 {\nim{#1},\ #2 (#3)}
\def\mypr   #1 #2 #3 {\pr{#1},\ #2 (#3)}

\def\myzpc  #1 #2 #3 {\zpc{#1},\ #2 (#3)}
\def\progtp #1 #2 #3 {Prog.~Theo.~Phys~{\bf#1},\ #2 (#3)}

\newcommand{\BABARPubYear}    {02}

\newcommand{\BABARPubNumber}  {010}

\newcommand{\SLACPubNumber} {9329}

\def\figurebox#1#2#3{%

    \def\arg{#3}%

    \ifx\arg\empty

    {\hfill\vbox{\hsize#2\hrule\hbox to #2{\vrule\hfill\vbox to #1{\hsize#2\vfill}\vrule}\hrule}\hfill}%

    \else

    {\hfill\epsfbox{#3}\hfill}%

    \fi}

\begin{document}
\preprint{SLAC-PUB-\SLACPubNumber} 
\begin{flushleft}
\babar-PUB-\BABARPubYear/\BABARPubNumber\\
SLAC-PUB-\SLACPubNumber\\
\end{flushleft}

\title{
{\large \bf A Study of the Rare Decays   \bdsordsspi\ and \bdsordssk } 
}
%
\author{B.~Aubert}
\author{D.~Boutigny}
\author{J.-M.~Gaillard}
\author{A.~Hicheur}
\author{Y.~Karyotakis}
\author{J.~P.~Lees}
\author{P.~Robbe}
\author{V.~Tisserand}
\author{A.~Zghiche}
\affiliation{Laboratoire de Physique des Particules, F-74941 Annecy-le-Vieux, France }
\author{A.~Palano}
\author{A.~Pompili}
\affiliation{Universit\`a di Bari, Dipartimento di Fisica and INFN, I-70126 Bari, Italy }
% ...
\collaboration{The \babar\ Collaboration}
\noaffiliation
 
%
\noaffiliation
\date{\today}

\begin{abstract}
We report  evidence for  the  decays \bdspi\  and \bdsk\ 
 and the results of a search for  \bdsspi\ and \bdssk\
in a sample  of \lumi\ \FourS\ decays into $B\overline{B}$  pairs collected 
with the \babar\ detector
at the PEP II asymmetric-energy $e^{+}e^{-}$ storage ring. 
We measure the branching fractions \brds\
and \brdsk.  We also set 90\% C.L. limits \brdsslim\ and \brdssklim. 
\end{abstract}


\maketitle

The measurement of the \CP-violating phase of the Cabibbo-Kobayashi-Maskawa  (CKM) matrix~\cite{CKM}
is an important part of the present scientific program in particle physics. 
\CP\ violation manifests itself as a non-zero area of the unitarity triangle~\cite{Jarlskog}.
While it is sufficient to measure one of the angles to demonstrate the
existence of \CP\ violation,
the unitarity triangle needs to be overconstrained by experimental measurements,
in order to demonstrate that the CKM mechanism is the correct 
explanation of this phenomenon. Several theoretically clean measurements
of the angle $\beta$ exist~\cite{sin2b}, but there is no such
measurement of the two other angles $\alpha$ and 
$\gamma $. A theoretically clean measurement of $\rm sin(2\beta+\gamma)$ can 
be obtained from the study
 of the time evolution for   $\Bz{\to} D^{(*)-} \pi^+$~\cite{chconj} decays, 
which are 
 already available in large samples at the \B\ factories, and for the corresponding
CKM-suppressed mode $\Bz{\to} D^{(*)+}\pi^-$~\cite{sin2bg}.
\begin{figure}[!htb]
\begin{center}
\includegraphics[width=\linewidth]{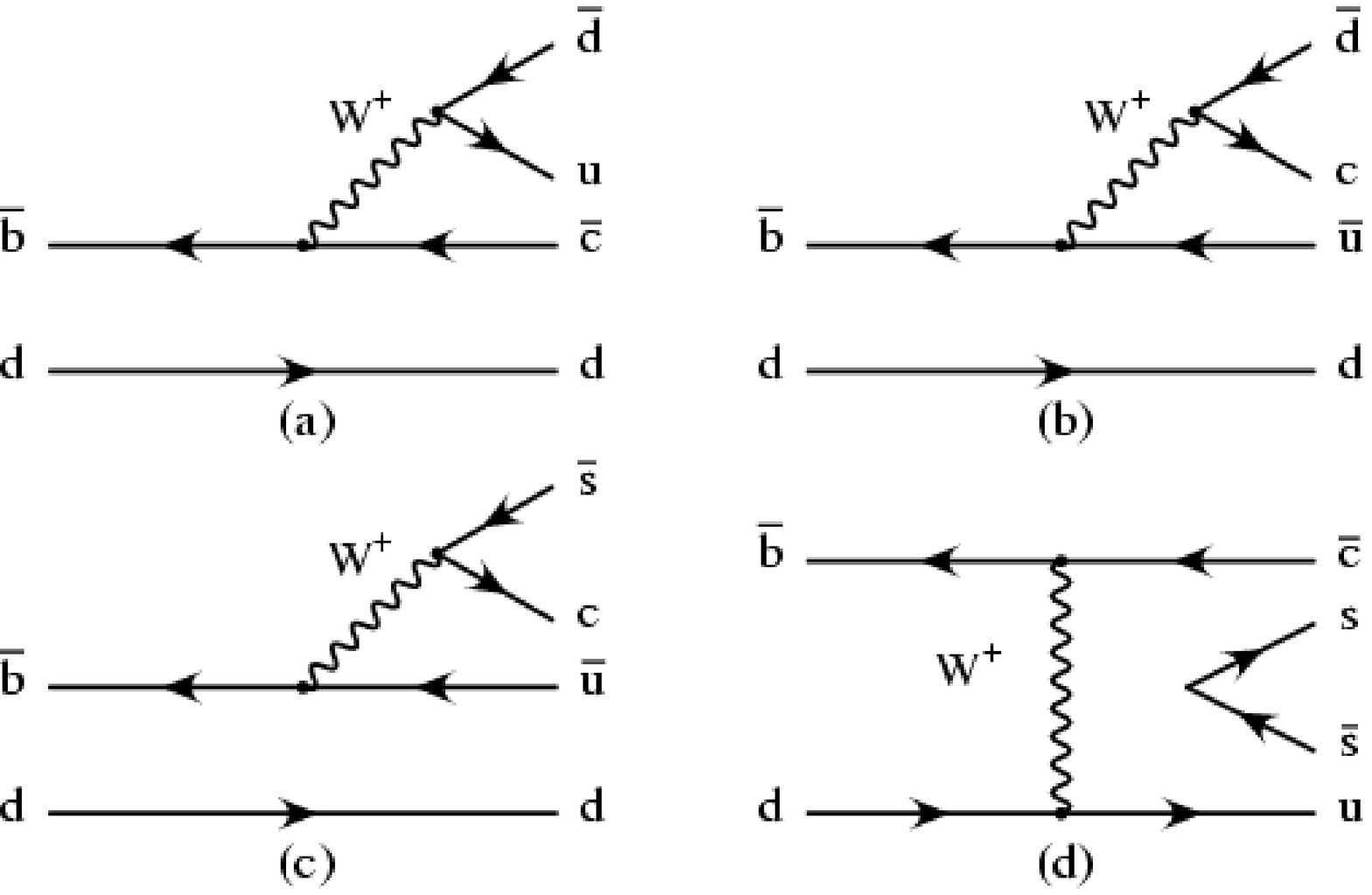}
\label{fig:feynman}
\caption{The Feynman diagrams for the decays a) $\Bz{\to} D^{(*)-} \pi^+$
, b) $\Bz{\to} D^{(*)+}\pi^-$, c) $\Bz{\to} D_s^{(*)+}\pi^-$, d)
$\Bz{\to} D_s^{(*)-}K^+$.}
\end{center}
\end{figure}

 This measurement 
requires a knowledge of the ratio 
of the decay amplitudes $R^{(*)}=|A(\Bz{\to} D^{(*)+}\pi^-)/A(\Bz{\to}
D^{(*)-}\pi^+)|$. 

Unfortunately a determination of $|A(\Bz{\to} D^{(*)+}\pi^-)|$ from a 
measurement of $\BR(\Bz{\to} D^{(*)+}\pi^-)$ is not
possible with the currently available data sample due to the presence
of the large background from $\Bzb{\to} D^{(*)+}\pi^-$.
However it has been suggested~\cite{sin2bg} that $R^{(*)}$ can be inferred from 
measurements of the ratios of the branching fractions 
$\BR(\bdsordsspi)/{\BR(\Bz{\to} D^{(*)-}\pi^+)} $ using SU(3) symmetry relation. 
The decays \bdsordsspi\ have also been proposed as a means for 
measuring   $ |V_{ub}/V_{cb}|$~\cite{roy}.

The decays \bdsordssk\ are a probe of the dynamics in $B$ decays because they
are expected to proceed  mainly via a
W-exchange diagram, not observed so far. In addition, these modes  can
be used to investigate the role of final state  rescattering, which 
can substantially  increase the expected rates~\cite{Wexch}.
Figure~1 shows the Feynman diagrams for the decays $\Bz{\to} D^{(*)-} 
\pi^+$, $\Bz{\to} D^{(*)+}\pi^-$, $\Bz{\to} D_s^{(*)+}\pi^-$ and 
$\Bz{\to} D_s^{(*)-}K^+$.

In this Letter we present  measurements of the
branching fractions for the decays  \bdsordsspi  and \bdsordssk.

The analysis uses  a sample  of \lumi\ \FourS\ decays into $B\overline{B}$
 pairs 
 collected  in the years 1999-2002 with the \babar\ detector
at the \pep2\ asymmetric-energy $B$-factory~\cite{pep}.
Since the \babar\ detector is described in detail
elsewhere~\cite{detector},  
only the components that are crucial to this analysis are
 summarized here. 
Charged particle tracking is provided by a five-layer silicon
vertex tracker (SVT) and a 40-layer drift chamber (DCH). 
For charged-particle identification, ionization energy loss ($dE/dx$) in
 the DCH and SVT, and Cherenkov radiation detected in a ring-imaging
 device are used. 
Photons are identified and measured using
the electromagnetic calorimeter, which comprises 6580 thallium-doped CsI
crystals. These systems are mounted inside a 1.5 T solenoidal
superconducting magnet. 
We use the GEANT~\cite{geant} software to simulate interactions of particles
traversing the \babar\ detector, taking into account the varying detector conditions and beam backgrounds.

We select events with a minimum of four reconstructed charged tracks and
a total measured energy greater than 4.5 GeV, determined
using all charged tracks and neutral clusters with energy above 30 MeV.
In order to reject continuum background, the ratio of the second
and zeroth order Fox-Wolfram moments~\cite{fox} 
must be less than 0.5.

So far, only upper limits have been reported for the modes studied here~\cite{prior}.
 Therefore the selection criteria are optimized
 to maximize the ratio of signal efficiency over the square-root of the
expected number of background events.

Candidates for \Ds\ mesons 
are reconstructed in the modes
 $\Ds {\to} \phi \pi^+$, $\KS K^+$ and $\Kstarzb\Kp$, 
with $\phi{\to} K^+K^-$, $\KS {\to} \pip \pim$, and $\Kstarzb{\to} K^-\pi^+$. 
 The $\KS$ candidates are reconstructed from two
oppositely-charged tracks with an invariant mass 
$493 < M_{\pip \pim} < 501\mevcc $. All other tracks are 
required  to originate from a vertex consistent with the
$\epem$ interaction point.
In order to identify charged kaons, two selections are used: a pion
veto with an efficiency of 95\% for kaons and  20\% 
for pions, and a tight kaon selection 
with an efficiency of 85\%  and 5\% pion misidentification probability.
Unless the tight selection is specified, the pion veto is always adopted. 
The $\phi$ candidates are reconstructed from two oppositely-charged 
kaons with an invariant mass $1009 < M_{\Kp \Km} < 1029\mevcc $. 
The $\Kstarzb$ candidates are constructed from  \Km\ and \pip\
candidates and are required to have an invariant mass in the range
$856 < M_{\Km \pip} < 936\mevcc $.
The polarization of the 
 \Kstarzb\ ($\phi$) mesons in the   
\Ds\ decays are also utilized to reject backgrounds through the use of the helicity angle
$\theta_H$, defined as the angle between one of the decay products  of
the \Kstarzb\ ($\phi$) 
and the direction of flight of the \Ds, in the  \Kstarzb\ ($\phi$)  rest
frame. 
Background events are distributed uniformly in $\cthel$ since they
originate from random combinations, while
signal events are distributed as $\ctwothel$.
The \Kstarzb\ candidates are therefore required to have $|\cthel|>0.4$, while
for the $\phi$ candidates we require $|\cthel|>0.5$. 
In order to reject background from $\Dp{\to}\KS\pip$ or $\Kstarzb\pip$,
the $\Kp$ in the reconstruction of $\Ds{\to}\KS\Kp$ or $\Kstarzb\Kp$ is
required to pass the tight 
kaon identification criteria introduced above.
Finally, the \Ds\ candidates are required to have an invariant mass within 10 \mevcc\ 
of the nominal value~\cite{PDG2002}.

We reconstruct  \Dss\ candidates in the mode  $\Dss{\to}\Ds\gamma$, by combining 
\Ds\ and photon candidates.
Photons that form a $\pi^0$ candidate, with $122< M_{\gamma\gamma}<147
 \mevcc$,
 in combination  with any other photon  with energy greater than 70 \mev are rejected.
 The mass difference between the \Dss\ and  the \Ds\ candidate is
 required to be within 14 \mevcc\ of 
the nominal value~\cite{PDG2002}. 

We combine \Dsos\ candidates with a track of opposite charge to form a $B$ candidate,
 and assign the candidate to the \bbardsordssk\ mode if the track satisfies the tight kaon 
selection and to the \bdsordsspi\ mode otherwise.
In order to reject events where the \Ds\
comes from a $B$ decay and the pion or kaon comes from the other $B$, 
we require the two decay products to
have a probability greater than 0.25\% of originating from a common
vertex. 

The remaining background is predominantly combinatorial in nature and
arises from continuum \qqbar\
production. 
This source is suppressed based on event topology. 
We compute the angle
($\theta_T$) between the thrust axis of 
 the  $B$ meson candidate and the thrust axis of  all  other particles in the event.
In the center-of-mass frame (c.m.), \BB\ pairs are
produced approximately at rest and  form a uniform $\cth$ distribution.
In contrast,
\qqbar\ pairs are produced back-to-back in the c.m. frame,
which results in a $|\cth|$
 distribution peaking at 1. 
Based on the background level of each mode, $|\cth|$ is required to
be smaller than a value that ranges between 0.7 and 0.8.
 We further suppress backgrounds using a Fisher discriminant $\cal{F}$ constructed from
the scalar sum of the c.m. momenta of all tracks and photons (excluding the $B$ candidate
decay products)  flowing into 9 concentric cones centered on the thrust axis of the $B$ candidate~\cite{twobody}. The more spherical the
event, the lower the value of ${\cal{F}}$. 
We require ${\cal{F}}$ to be smaller than a threshold that varies from
  0.04 to 0.2 depending on the background level.
 
\begin{figure}[!htb]
\begin{center}
\includegraphics[width=\linewidth]{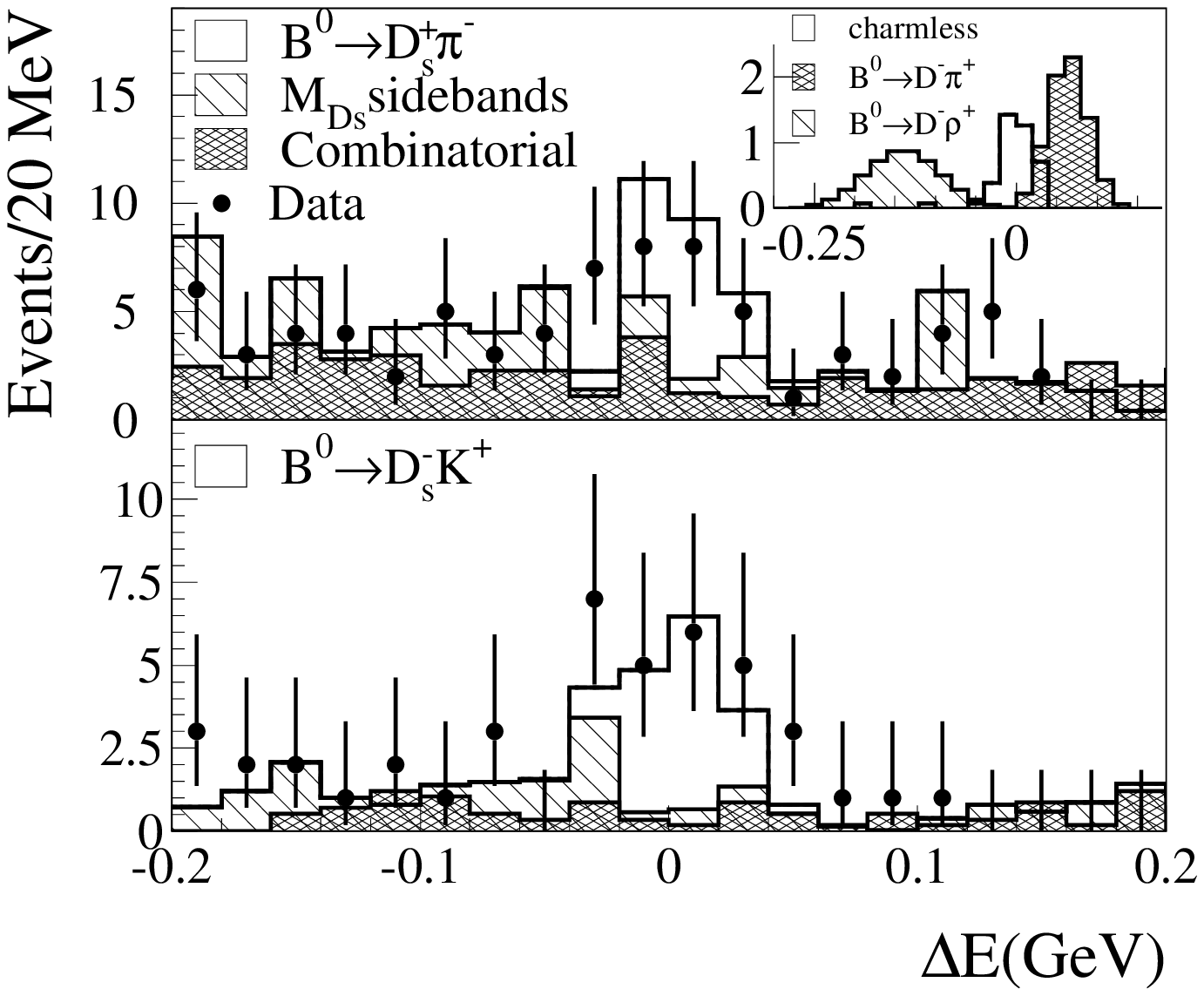}
\label{fig:datade}
\caption{ The \de\ distribution for \bdspi\ (top) and \bdsk\ (bottom) candidates in data compared with the distributions of
   the combinatorial background, estimated from the \mes\ sideband,
  the cross-contamination, estimated from the \mds\
  sidebands, and the simulation of the signal, normalized to the observed yield.
The insert shows  the  \de\ distribution of the separate
  contributions to the cross contamination to the \bdspi\ signal as predicted by 
  simulation. The reflection backgrounds are normalized to the known
  branching fractions~\cite{PDG2002}, while the normalization of the
  charmless background
is arbitrary.
}

\end{center}
\end{figure}

We extract the signal using the kinematic variables
$\mes= \sqrt{E_{\rm b}^{*2} - (\sum_i {\bf p}^*_i)^2}$
and $\de= \sum_i\sqrt{m_i^2 + {\bf p}_i^{*2}}- E_{\rm b}^*$,
where $E_{\rm b}^*$ is the beam energy in the c.m. frame,
${\bf p}_i^*$ is the c.m.~momentum of daughter particle $i$ of the
$B$ meson candidate, and $m_i$ is the mass hypothesis for particle $i$. 
For signal events, $m_{\rm ES}$ peaks at the $B$ meson mass with 
a resolution of about 2.5 MeV$/c^2$ and $\Delta E$ peaks near zero,  
indicating that the candidate system of particles has total energy consistent with
the beam energy in the c.m.~frame.
The 
  \de\ signal band is defined by $|\de-5|<36 \mev$ and within the band
we define the events with $\mes>5.27$\gevcc  as the signal candidates.

After the aforementioned selection, three classes of backgrounds remain.
First, the amount of {\it{combinatorial background}} in the signal region is 
estimated from the sideband of 
the \mes\ distribution which is
described by a threshold function 
$\frac{dN}{dx}=x\sqrt{1-x^{2}/E_b^{*2}}\exp\left[-\xi\left(1-x^{2}/E_b^{*2}\right)\right]$, 
characterized by the shape parameter $\xi$~\cite{argus}. 

Second,  $B$ meson decays such as  $\Bzb{\to}\Dp\pim, \rho^-$ with
$\Dp{\to}\KS\pip$ or $\Kstarzb\pip$ can 
constitute a background for  the \bdspi\ mode if
 the pion in the $D$ decay is misidentified as a kaon ({\it{reflection
background}}). These backgrounds have the same \mes\ distributions 
as the signal but different distributions in \de.
The corresponding backgrounds for the
\bdsk\ mode ($\Bz{\to}\Dm\Kp,\Kstarp$) have a branching fraction ten times smaller.

Finally, rare $B$  decays into the same final state, such as
$\Bz{\to} \Kbar^{(*)0}\Kp\pim$ or $\Kbar^{(*)0}\Kp\Km$
({\it{charmless background}}),
have the same \mes\ and \de\ distributions as the \bdspi\ or \bdsk\ signal.
Figure~2 shows the \de\ distribution for the \bdspi\ and \bdsk\ signal and
for various sources of background.
The branching fraction of the charmless background is not well measured;
 therefore we need to  estimate the sum of the reflection and charmless background (referred
to as {\it{cross-contamination}}) directly with data. This is possible
because both of these background sources have a flat distribution in
the \Ds\ candidate  mass (\mds) while the signal has a Gaussian distribution.

Possible contamination 
from $B{\to} D_s^{(*)} X$ decays is determined with simulation and found to be negligible. 
The cross-contamination for the decays \bdsspi\ and \bdssk\ is dominated
by the reflection background, which we  estimate from simulation.
Cross-feed between \bdsordsspi\ and \bdsordssk\ modes is estimated
 to be less than 1\%.

\begin{figure}[!htb]
\begin{center}

\includegraphics[width=\linewidth]{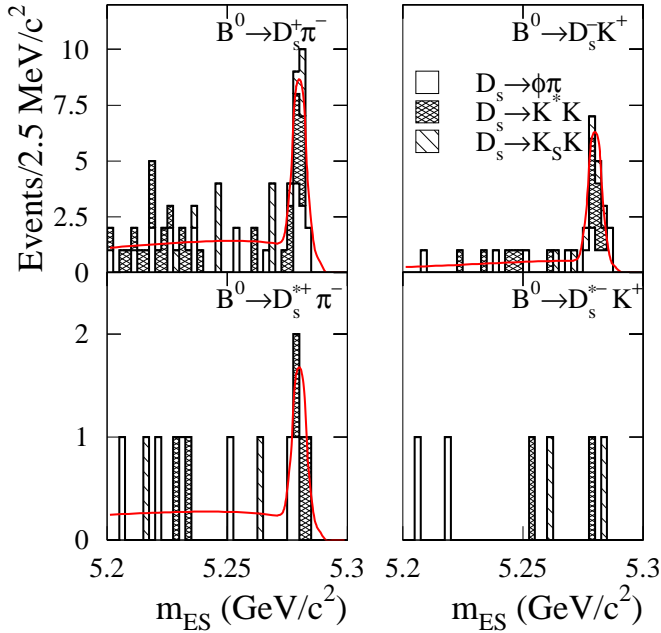}
\label{fig:datamb}

\caption{The \mes\ distributions for the  \bdspi\ (top left), \bdsk\
  (top right), \bdsspi\ (bottom left), and \bdssk\
  (bottom right)
candidates within the \de\ band in data  after all selection requirements. 
The fits used to obtain the signal yield are described in the text. The
  contribution from 
each \Ds\ mode is shown separately. 
}
\end{center}
\end{figure}


Figure~3 shows the \mes\ distribution in the \de\ signal band
 for each of the modes. 
We perform an unbinned
maximum-likelihood fit  to each
 \mes\ distribution with a threshold function to characterize the
 combinatorial background and a Gaussian distribution to describe the sum of 
the signal and cross-contamination contributions. 
 The mean and the width of the Gaussian distribution are fixed to the 
values obtained in a copious $\Bz{\to} D^{(*)-} \pip$ control sample.
 For the \bdspi\ and \bdsk\ analyses, we obtain the threshold parameter $\xi$  from a
 fit to the distributions of \mes\ in data, after loosening the \mds\ and
 \de\ requirements. In the case of \bdsspi\ and \bdssk, due to the low 
 background level,  we use simulated events to estimate $\xi$.

No fit is performed with the \bdssk\ sample due to the small number of events.
Whenever there are enough events, we   fit each \Ds\ decay
mode separately, as well as the combination of all modes.
The cross-contamination is estimated by performing the same fit on the
events in the data \mds\ sidebands ($4\sigma<|\mds-1968.6\mevcc|<8\sigma$,
where the resolution is $\sigma=5\mevcc$).
The number of observed events, the background expectations, and the 
reconstruction efficiencies  estimated with simulated events are
summarized in Table~\ref{tab:fit}.


\begin{table*}[t!]
\caption{The number of signal candidates ($N_{\rm{sigbox}}$), 
the  Gaussian yield ($N_{\rm {gaus}}$) and  the combinatorial background 
($N_{\rm{comb}}$) extracted from the likelihood fit, the cross-contamination ($N_{\rm{cross}}$),
the reconstruction efficiency ($\varepsilon$), 
 the probability ($P_{\rm{bckg}}$) of the data being consistent with 
the background fluctuating up to the level of the data in the 
absence of signal, the measured branching fraction (\BR
), and the 90\% confidence-level upper limit. 
$N_{\rm {gaus}}$, $N_{\rm{comb}}$, and \BR\  are not available for modes with too few events.
$N_{\rm{cross}}$ is not reported if no event is found in the \Ds\ mass sideband.
}

\begin{center}
\mbox{ \scriptsize
\begin{tabular}{l|c|c|c|c|c||c|c|c} \hline \hline
$B$ mode& $N_{\rm{sigbox}}$ & $N_{\rm {gaus}}$  & $N_{\rm{comb}}$  & $N_{\rm{cross}}$ &
$\varepsilon$(\%)&  $P_{\rm{bckg}}$ & \BR ($10^{-5}$)  & 90\%
C.L. \\
& & & & & & & & ($10^{-5}$)\\ \hline
\bdspi & & & & & & & & \\
\ \ $\Ds{\to}\phi\pip$    &9 &$8.0\pm3.0$  &$2.1\pm0.7$ &$<0.7$ &16.9  &$1.4\times10^{-3}$ & $3.1 \pm 1.2$ & -\\
\ \ $\Ds{\to}\Kstarzb\Kp$ &12 &$9.2\pm3.4$ &$3.8\pm1.0$ &$2.9\pm1.8$ & 9.6  &$2.3\times10^{-2}$ & $3.5 \pm 1.9$ & -\\
\ \ $\Ds{\to}\KS\Kp$      &5 &$4.2\pm2.2$  &$1.9\pm0.6$ &$1.2\pm1.4$ &12.3  &$8.3\times10^{-2}$ &$2.4 \pm 1.8$  & -\\ 
\ \ all                   &26 &$21.4\pm 5.1$ &$7.8\pm1.7$ &$3.7\pm2.4$       &N/A   &$9.5\times10^{-4}$ & $3.2\pm 0.9\pm 1.0$&-\\ \hline
\bdsspi & & & & & & & & \\
\ \ $\Ds{\to}\phi\pip$    &2 &- &$0.6\pm0.3$ &$<0.14$ &7.8  &- &- &- \\
\ \ $\Ds{\to}\Kstarzb\Kp$ &3 &$2.8_{-1.8}^{+2.7}$ & $0.4\pm0.3$ & $0.3\pm0.2$ & 3.3  &$3.9\times 10^{-2}$  &$4.3_{-3.1}^{+4.7}$  &$<12$  \\
\ \ $\Ds{\to}\KS\Kp$      &0 &- &$0.4\pm0.3$ & $<0.14$ & 5.1  &- &-&-\\ 
\ \ all                   &5 &$4.4_{-2.8}^{+2.7}$ &$1.2\pm0.4$  &$0.3\pm0.2$ &N/A & $2.3\times 10^{-2}$ & $1.9_{-1.3}^{+1.2}\pm0.5$  & $<4.1$ \\ \hline
\bdsk  & & & & & & & & \\
\ \ $\Ds{\to}\phi\pip$    &7 &$5.8\pm2.6$ &$1.3\pm0.7$ &$1.1\pm1.2$ &13.0  &$4.5\times10^{-2}$ &$2.4\pm1.3$ &- \\
\ \ $\Ds{\to}\Kstarzb\Kp$ &8 &$7.3\pm2.9$ &$1.7\pm0.7$ &$<0.7$ &7.8  &$1.9\times10^{-3}$ &$5.0\pm2.0$  &- \\
\ \ $\Ds{\to}\KS\Kp$      &4 &$3.7\pm2.0$ &$0.6\pm0.4$ &$1.3\pm1.0$ &9.2  &$1.7\times10^{-2}$ &$2.5\pm2.1$ &- \\ 
\ \ all                   &19 &$16.7\pm4.3$ &$3.5\pm1.3$ &$2.7\pm1.9$ &N/A &$5.0\times10^{-4}$  & $3.2\pm1.0\pm1.0$ & -\\ \hline
\bdssk & & & & & & & & \\
\ \ $\Ds{\to}\phi\pip$    &0 &- &$0.8\pm0.6$ &$<0.14$&5.3 &- &- &- \\
\ \ $\Ds{\to}\Kstarzb\Kp$ &1 &- &$0.4\pm0.4$ &$<0.14$ &2.7 &- &- &- \\
\ \ $\Ds{\to}\KS\Kp$      &1 &- &$0.4\pm0.4$ &$<0.14$  &4.3 &- &- &- \\ 
\ \ all                   &2 &- &$1.6\pm0.8$ &$<0.14$  &N/A &0.48 &- &$<2.5$ \\ \hline
\end{tabular}}
\end{center}
\label{tab:fit}
\end{table*}

In the \bdspi (\bdsk) mode the fit yields a
Gaussian contribution of $21.4\pm 5.1$ ($16.7\pm 4.3$) events and a
combinatorial background of $7.8\pm 1.7$ ($3.5\pm 1.3$) events. The
cross-contamination is 
estimated to be $3.7\pm 2.4$ ($2.7\pm 1.9$) events. The probability of
the background to fluctuate to the observed number of events, taking
into account both Poisson statistics and uncertainties in the
background estimates, is $9.5\times 10^{-4}$ ($5.0\times
10^{-4}$). For a Gaussian distribution this would correspond to
$3.3\sigma$ ($3.5\sigma$).
Given the estimated reconstruction efficiencies we measure
$\BR(\bdspi)=(3.2\pm 0.9)\times 10^{-5}$ ($\BR(\bdsk)=(3.2\pm 1.0)\times 10^{-5}$),
where the quoted error is statistical only.
We also set the  90\% C.L. limits  \brdsslim\ and \brdssklim.

The systematic errors are dominated by the 25\% relative uncertainty for
 \BR (\Ds$\rightarrow\phi\pip$). The uncertainties on the knowledge of
 the background come from uncertainties in the $\xi$ parameter, for the
 combinatorial background, and from  the limited number of events in the
 \mds\ sidebands for the cross-contamination. They amount to 14\%, 16\%, 7\%, and 36\% of the measured
 branching fractions in the  \bdspi , \bdsk, \bdsspi, and \bdssk\ modes, respectively.
The rest of the systematic errors, which include the uncertainty on tracking, \KS\ reconstruction, and charged-kaon identification
 efficiencies, range between   11\% and 14\% depending on the mode.

 
In conclusion, 
we  report a 3.3$\sigma$ signal for the \btou\ transition 
 \bdspi\ and a 3.5$\sigma$ signal for the decay \bdsk, and measure
\begin{center}
 \brds , \\
 \brdsk.
\end{center}
Since the dominant uncertainty 
comes from the knowledge of the \Ds\ branching fractions we also compute
$\BR( \bdspi )\times \BR(\Ds\rightarrow\phi\pip)=(1.13\pm0.33\pm0.21)
\times 10^{-6}$
and $\BR( \bdsk )\times \BR(D_s^-\rightarrow\phi\pim)=(1.16\pm0.36\pm0.24)
\times 10^{-6}$.
The search for \bdsspi\ and \bdssk\  yields the 90\% C.L. upper limits
\begin{center}
 \brdsslim ,  \\
 \brdssklim.
\end{center}

We are grateful for the excellent luminosity and machine conditions
provided by our \pep2\ colleagues, 
and for the substantial dedicated effort from
the computing organizations that support \babar.
The collaborating institutions wish to thank 
SLAC for its support and kind hospitality. 
This work is supported by
DOE
and NSF (USA),
NSERC (Canada),
IHEP (China),
CEA and
CNRS-IN2P3
(France),
BMBF and DFG
(Germany),
INFN (Italy),
NFR (Norway),
MIST (Russia), and
PPARC (United Kingdom). 
Individuals have received support from the 
A.~P.~Sloan Foundation, 
Research Corporation,
and Alexander von Humboldt Foundation.


\begin{thebibliography}{99}
\bibitem{chconj}
Charge conjugation is implied throughout this letter, unless 
explicitly stated.
\end{thebibliography}
\end{document}